\documentclass[Afour,sageapa,times]{sagej}

\usepackage{moreverb,url}

\usepackage{graphicx}
\usepackage{multicol}
\usepackage{subcaption}
\usepackage{amssymb}
\usepackage{pifont}
\newcommand{\xmark}{\ding{55}}%

\usepackage{enumitem}
\newcommand\BibTeX{{\rmfamily B\kern-.05em \textsc{i\kern-.025em b}\kern-.08em
T\kern-.1667em\lower.7ex\hbox{E}\kern-.125emX}}

\begin{document}

\runninghead{R. Kelter}

\title{How to choose between different Bayesian posterior indices for hypothesis testing in practice}

\author{Riko Kelter\affilnum{1}}

\affiliation{\affilnum{1}Department of Mathematics, University of Siegen}

\corrauth{Riko Kelter, Department of Mathematics, University of Siegen,
Walter-Flex-Street 3,
57072 Siegen,
Germany}

\email{riko.kelter@uni-siegen.de}

\begin{abstract}
Hypothesis testing is an essential statistical method in psychology and the cognitive sciences. The problems of traditional null hypothesis significance testing (NHST) have been discussed widely, and among the proposed solutions to the replication problems caused by the inappropriate use of significance tests and $p$-values is a shift towards Bayesian data analysis. However, Bayesian hypothesis testing is concerned with various posterior indices for significance and the size of an effect. This complicates Bayesian hypothesis testing in practice, as the availability of multiple Bayesian alternatives to the traditional $p$-value causes confusion which one to select and why. In this paper, we compare various Bayesian posterior indices which have been proposed in the literature and discuss their benefits and limitations. Our comparison shows that conceptually not all proposed Bayesian alternatives to NHST and $p$-values are beneficial, and the usefulness of some indices strongly depends on the study design and research goal. However, our comparison also reveals that there exist at least two candidates among the available Bayesian posterior indices which have appealing theoretical properties and are, to our best knowledge, widely underused among psychologists.
\end{abstract}


\maketitle

\renewcommand*{\thefootnote}{\arabic{footnote}}
\section{Introduction}
Hypothesis testing is a standard statistical method widely adopted in psychological research. Historically, the last century has included the advent of various proposals to test research hypotheses statistically \shortcite{Howie2002,Mayo2018}. The most well-known approaches include the theory of significance testing of British statistician Ronald Fisher \cite{Fisher1925SMRWI} and the anglo-polish Neyman-Pearson theory of hypothesis testing \shortcite{Neyman1933}. While the former interprets the $p$-values as a continuous quantitative measure against the null hypothesis, the latter is targeted at bounding the type I error rate to a specific test level $\alpha$ while simultaneously minimising the type II error rate. In the Neyman-Pearson theory $p$-values are interpreted as binary values which either show the significance of a result or do not. Conceptually, however, both approaches are located in the frequentist school of thought \cite{Howie2002,Nuzzo2014}.

The problems of frequentist hypothesis testing, in general, have been detailed and discussed widely in recent years \shortcite{Colquhoun2014,wasserstein2016,Colquhoun2017,benjaminRedefineStatisticalSignificance,Wasserstein2019,Benjamin2019}. One of the largest problems of frequentist hypothesis testing is its conflict with the likelihood principle, which has a nearly axiomatic role in statistical science \shortcite{Cox1958,Birnbaum1962,Basu1975}. Among the consequences of this conflict are the inability to use optional stopping (that is, stop recruiting study participants when the data already show overwhelming evidence and report the results) \shortcite{Edwards1963}, problems with the interpretation of censored data which often are observed in psychological research \cite{Berger1988a} and the dependency of the results on the researcher's intentions \cite{Kruschke2018}.

The range of proposals which have been made to solve the problems of NHST and $p$-values is huge. These proposals include stricter thresholds for stating statistical significance \shortcite{benjaminRedefineStatisticalSignificance}, recommendations on how to improve the use of $p$-values in practice \cite{Benjamin2019} and more serious methodological shifts \shortcite{Rouder2014,Morey2016c,Morey2016,Kruschke2018}. Among the latter, the shift towards Bayesian statistics as an attractive alternative is proposed often, in particular, to improve the situation in psychology \shortcite{Wagenmakers2010,Dienes2011,Dienes2014}.

Even when putting the problems of NHST and $p$-values aside for a moment and considering a wider timeframe, Bayesian statistics' popularity in psychology has grown. Van de Schoot et al. \citeyear{VanDeSchoot2017} conducted a large systematic review of $n=1579$ Bayesian psychologic articles published in the years between $1990$ and $2015$. According to them, Bayesian statistics has become increasingly popular among psychologists and ``Bayesian statistics is used in a variety of contexts across subfields of psychology and related disciplines.'' \shortcite[p.~1]{VanDeSchoot2017}. Furthermore, they noted that
\begin{quote}
    ``There are many different reasons why one might choose to use Bayes (e.g., the use of priors, estimating otherwise intractable models, modeling uncertainty, etc.). We found in this review that the use of Bayes has increased and broadened in the sense that this methodology can be used in a flexible manner to tackle many different forms of questions.''\newline
    \shortcite[p.~1]{VanDeSchoot2017}
\end{quote}
For a good review of why and how Bayesian statistics improves the reliability of psychological science see Dienes \citeyear{Dienes2018}. For a less optimistic perspective and criticism on Bayes factor based hypothesis testing in psychological research see Tendeiro and Kiers \cite{Tendeiro2019}. Maybe the most important reason why Bayesian statistics (including Bayesian hypothesis testing) may be preferable to classic significance testing is its accordance with the likelihood principle. This implies that results do not depend on researchers' intentions, optional stopping is allowed and interpretation of censored data is simplified \shortcite{Berger1988a,Rouder2014,Wagenmakers2016, Kruschke2018}.

In Bayesian data analysis, inference is concerned with the posterior distribution $p(\theta|x)$ after observing the data $x$. To obtain this posterior distribution, a prior distribution $p(\theta)$ is assumed for the unknown parameter $\theta$, and the prior distribution is updated via the model likelihood $p(x|\theta)$ to the posterior distribution:
\begin{align}
    p(\theta|x)\propto p(x|\theta)p(\theta)    
\end{align}
Here, $\propto$ means proportional to and in practice it suffices to obtain a quantity proportional to the posterior $p(\theta|x)$ due to modern sampling algorithms which are used to obtain the posterior $p(\theta|x)$ in practice \shortcite{Robert2004,Carpenter2017,McElreath2020}.

However, Bayesian statistics also has limitations and some aspects pose serious challenges to practitioners. For a thorough review of these problems we refer to Tendeiro \& Kiers \citeyear{Tendeiro2019}, and here we list only three important issues.

First, among the challenges of applying Bayesian data analysis in practice is the selection of the prior distribution $p(\theta)$. Extreme choices of prior distributions can distort the obtained results, yield unrealistic parameter estimates, or even influence the obtained posterior estimates into a desired direction \shortcite{Kelter2020BayesianPosteriorIndices,Gabry2020RstanarmPriorsVignette}.

Second, obtaining the posterior distribution $p(\theta|x)$ can become computationally expensive, in particular when the statistical model is high-dimensional and massive amounts of data are observed \shortcite{Magnusson2019}. However, this limitation is a less severe problem due to the advent of modern sampling algorithms like Markov-Chain-Monte-Carlo (MCMC) \shortcite{Robert2008,Diaconis2009} and, in particular, Hamiltonian Monte Carlo (HMC) \shortcite{Neal2011,Carpenter2017,Kelter2020BMCJasp}.

Third, concerning Bayesian hypothesis testing as a replacement or alternative to NHST and $p$-values, there exist a variety of posterior indices which have been proposed in the statistics literature \shortcite{Makowski2019,Kelter2020BayesianPosteriorIndices}. Conceptually, these posterior indices are all based on the posterior distribution $p(\theta|x)$ im some form, and are employed to test a null hypothesis like $H_0:\theta=\theta_0$ against the alternative $H_1:\theta \neq \theta_0$ (or against a one-sided alternative $H_1:\theta <0$ or $H_1:\theta >0$). The mathematical theory behind each of the proposed posterior indices differs substantially, and examples include the Bayes factor \shortcite{jeffreys1961,KassRaftery1995,Morey2016,Held2018a}, the region of practical equivalence (ROPE) \shortcite{Westlake1976,Kirkwood1981,Kruschke2018,Kruschke2018a}, the MAP-based $p$-value \shortcite{Mills2017}, the probability of direction (PD) \shortcite{Makowski2019} and the Full Bayesian Significance Test (FBST) and the $e$-value \shortcite{Pereira1999,Pereira2008,Stern2020}. While there are some results available which compare the available posterior indices for settings like linear regression \shortcite{Makowski2019} or parametric two-sample tests \cite{Kelter2020BayesianPosteriorIndices}, in general, the suitability of a given posterior index for testing a hypothesis depends both on the study design and research goal, see Kelter \citeyear{Kelter2020BayesianPosteriorIndices}.

In this paper, we focus on the third problem and provide a conceptual comparison of the most widely used Bayesian posterior indices for hypothesis testing, discuss the benefits and limitations of each index and give guidance for when to apply which posterior index. We do not limit the discussion to a specific method of choice like regression models or the analysis of variance, because most of the proposed indices are widely applicable and the limitations and benefits of each posterior index apply independently of the statistical model considered. However, to foster understanding and highlight the practical relevance we use a running example and apply each index discussed to it after giving a theoretical review of the index. This way, readers can get familiar with how the different posterior indices work and get a feeling of how they differ in a practical setting. As a running example, we use the Bayesian two-sample t-test of Rouder et al. \citeyear{Rouder2009}, which provides the posterior distribution $p(\delta|x)$ of the effect size $\delta$ \cite{cohen_statistical_1988} after observing the data $x$.

The structure of the paper is as follows: First, the running example is presented which builds the foundation for the application of any posterior index in the second step. Second, various available posterior indices are introduced, their benefits and limitations are discussed and they are applied to the running example. These include the Bayes factor \shortcite{jeffreys1961,KassRaftery1995,Morey2016,Held2018a}, the region of practical equivalence (ROPE) \shortcite{Westlake1976,Kirkwood1981,Kruschke2018,Kruschke2018a}, the MAP-based $p$-value \shortcite{Mills2017}, the probability of direction (PD) \shortcite{Makowski2019} and the Full Bayesian Significance Test (FBST) and the $e$-value \shortcite{Pereira1999,Pereira2008,Stern2020}. Third, a detailed discussion and comparison are provided. Finally, we conclude by giving guidance on how to select between the available posterior indices in practice, where it is shown that the choice of a posterior index depends on both the study design and research goal.

\section{Running example}
As mentioned above, the Bayesian two-sample t-test of Rouder et al. \citeyear{Rouder2009} is used as a running example in this paper. The test assumes normally distributed data in both groups and employs a Cauchy prior $C(0,\gamma)$ on the effect size $\delta$, that is $p(\delta)=C(0,\gamma)$. Using Markov-Chain-Monte-Carlo (MCMC) sampling, the posterior distribution $p(\delta|x)$ can be obtained, which then subsequently can be used for Bayesian hypothesis testing via a posterior index. Widely used choices of $\gamma$ include $\sqrt{2}/2$, $1$ and $\sqrt{2}$, which correspond to a medium, wide and ultrawide prior. Here, the wide prior $p(\theta)=C(0,1)$ is used, which places itself between the two other more extreme options. Note that we do not discuss prior selection extensively here, but in general, the two-sample t-test of Rouder et al. \citeyear{Rouder2009} is quite robust to the prior selection, for details see Kelter \citeyear{Kelter2020BayesianPosteriorIndices}.

In the example, $n=50$ observations are used in both groups, and data in the first group is simulated from a $\mathcal{N}(2.51,1.81)$ normal distribution, and data in the second group is simulated from a $\mathcal{N}(1.72,1.51)$ normal distribution. As a consequence, the true effect size $\delta_t$ according to Cohen \citeyear{cohen_statistical_1988} is given as
\begin{align}
    \delta_t:=\frac{(2.71-1.71)}{\sqrt{(1.81^2+1.51^2)/2}}\approx 0.60
\end{align}
which equals the existence of a medium effect. Figure \ref{fig:1} shows the posterior distribution $p(\delta|x)$ of the effect size $\delta$ based on the wide Cauchy prior $C(0,1)$ after observing the simulated data of both groups. Clearly, the distribution is shifted away from $\delta=0$, but it also does not centre strongly around the true value $\delta_t=0.60$. This can be attributed both to the amount of data observed and to the randomness in simulation. The vertical blue line shows the resulting posterior mean of $\delta$, which equals $\delta=0.42$.

\begin{figure}[!h]
\centering
\includegraphics[width=0.5\textwidth]{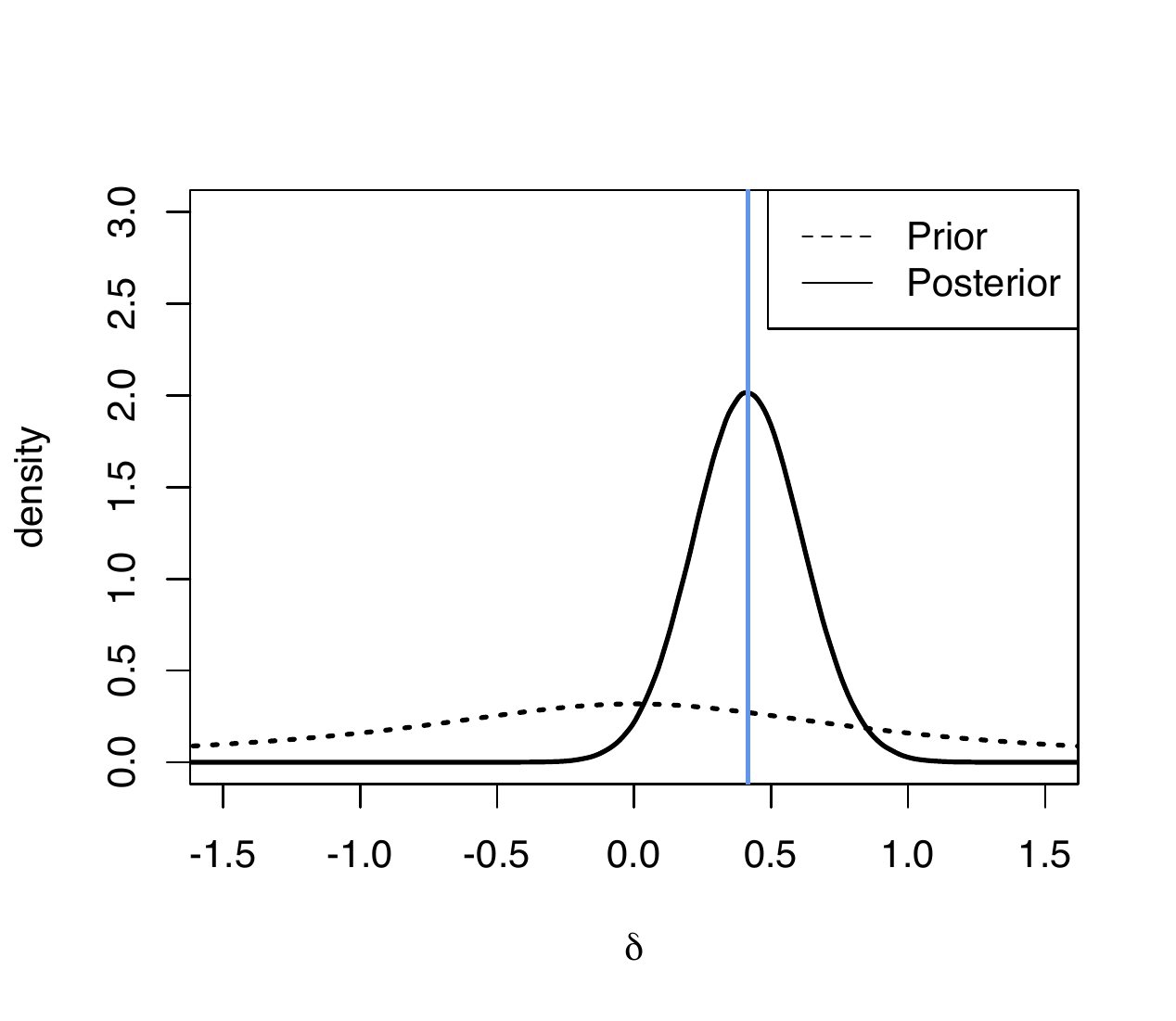}
\caption{Prior-posterior plot for the effect size of the two-sample Bayesian t-test in the running example. Dashed line shows the prior distribution $p(\delta)=C(0,1)$, which is a wide Cauchy prior distribution. Solid line shows the posterior distribution $p(\delta|x)$ after observing the data $x$ in both groups. The vertical blue line is located at the posterior mean of $\delta$, which is $0.31$.}
\label{fig:1}
\end{figure}

However, by now no posterior index has been employed to test the null hypothesis $H_0:\delta=0$ against the alternative $H_1:\delta \neq 0$. Based on the true effect size $\delta_t= 0.60$, any Bayesian posterior index should reject the null hypothesis $H_0:\delta=0$ in favour of the alternative $H_1:\delta \neq 0$, or even accept the alternative hypothesis $H_1:\delta \neq 0$ if possible.

All analyses and results including the figures shown in this manuscript can be reproduced via the replication script provided at the Open Science Foundation under \url{https://osf.io/xnfb2/}.

\section{Bayesian posterior indices}
This section discusses the various available posterior indices for testing a hypothesis in the Bayesian paradigm. The underlying theory, benefits and limitations of each index are detailed and afterwards, the index is applied to the running example shown in figure \ref{fig:1}.

\subsection{The Bayes factor}
The Bayes factor (BF) is one of the oldest and still widely used indices for testing a hypothesis in the Bayesian approach. It was invented by Sir Harold Jeffreys \citeyear{Jeffreys1931,jeffreys1961} and compares the predictive ability of two competing models corresponding to each of the hypotheses $H_0$ and $H_1$ under consideration. The Bayes factor $BF_{01}$ can be interpreted as the predictive updating factor which measures the change in relative beliefs about both hypotheses $H_0$ and $H_1$ after observing the data $x$:
\begin{align}\label{eq:bf}
    \underbrace{\frac{\mathbb{P}(H_0|x)}{\mathbb{P}(H_1|x)}}_{\text{Posterior odds}}    =\underbrace{\frac{p(x|H_0)}{p(x|H_1)}}_{BF_{01}(x)}\cdot \underbrace{\frac{\mathbb{P}(H_0)}{\mathbb{P}(H_1)}}_{\text{Prior odds}}
\end{align}
Phrased differently, the Bayes factor $BF_{01}$ is the ratio of the two marginal likelihoods $p(x|H_0)$ and $p(x|H_1)$, each of which is calculated by integrating out the respective model parameters according to their prior distribution. Generally, the calculation of these marginals can be complex in realistic models \cite{Rubin1984,Held2014}.

However, there exist sophisticated numerical techniques to obtain a Bayes factor in practice even for models which are analytically difficult to handle. Examples are \textit{bridge sampling} \shortcite{Gronau2017,Gronau2019a} or the \textit{Savage-Dickey density ratio} \shortcite{Dickey1970,Verdinelli1995}, which have become increasingly popular among psychologists \shortcite{Wagenmakers2010}.

Considering the setting of the Bayesian two-sample t-test in the running example, the Bayes factor is employed to test the null hypothesis $H_0:\delta=0$ of no effect against a the two-sided alternative $H_1:\delta \neq 0$ (one-sided alternatives $H_1:\delta >0$ or $H_1:\delta <0$ are of course also possible, but here we focus on a two-sided alternative). Here, $\delta=(\mu_1-\mu_2)/\sigma$ is the well-known effect size of \cite[p.~20]{cohen_statistical_1988}, under the assumption of two independent samples and identical standard deviation $\sigma$ in each group.

An appealing property of the Bayes factor is its ability to state evidence both for the null hypothesis $H_0$ and the alternative $H_1$. In contrast, traditional $p$-values can only signal evidence against the null hypothesis $H_0$. However, in practice, researchers are often not only interested in rejecting a null hypothesis $H_0$. When rejecting such a hypothesis, the natural question which follows is: How large is the true effect size $\delta_t$, when the evidence against $H_0:\delta=0$ is substantial so that one would reject the hypothesis that no effect exists? While a $p$-value was not designed to answer such a question, the Bayes factor $BF_{01}$ quantifies the necessary change in beliefs and can provide this information. If $BF_{01}>1$, the data indicate a necessary change in beliefs towards the null hypothesis $H_0$. If on the other hand $BF_{01}<1$, the data show evidence for the alternative $H_1$. Notice that from equation (\ref{eq:bf}), the relationship $BF_{01}=1/BF_{10}$ holds.

While $BF_{01}=1$ is a natural threshold to separate between evidence for the null and the alternative hypothesis, it is more difficult to interpret different magnitudes of the Bayes factor. Over the last century, various scales have been proposed which translate given magnitudes of the Bayes factor into evidence. The oldest one is the scale of Jeffreys himself \cite{Jeffreys1931,jeffreys1961}, and other proposals have been made by Kass \& Raftery \citeyear{KassRaftery1995}, Goodman \citeyear{Goodman1999}, Lee and Wagenmakers \citeyear{LeeWagenmakers2013}, Held and Ott \citeyear{Held2016} and van Doorn et al. \citeyear{VanDoorn2019}. Table 1 gives an overview about the different scales and shows that these differ only slightly.\footnote{Notice that Jeffreys \citeyear{jeffreys1961} used the cut points $(1/\sqrt{10})^a$ with $a=1,2,3,4$, and Goodman \citeyear{Goodman1999} used $1/5$, $1/10$, $1/20$ and $1/100$ for \textit{weak}, \textit{moderate}, \textit{moderate to strong} and \textit{strong to very strong}, which have been aligned with the cut points in the left column of table 1 to simplify comparison of the different scales.} As a consequence, translating a given Bayes factor into a statement about the provided evidence is, therefore, a less severe problem in practice.

\begin{table*}[bt]
\caption{Categorization of Bayes factors $BF_{01}\leq 1$ into evidence against $H_0$}
\begin{tabular}{lcccc}
\toprule
Bayes factor & \multicolumn{4}{c}{Strength of evidence against $H_0$}\\
& \cite{jeffreys1961} & \cite{Goodman1999} & \cite{Held2016} & \cite{LeeWagenmakers2013}\\
\midrule
$1$ to $1/3$ & Bare mention & & Weak & Anecdotal \\
$1/3$ to $1/10$ & Substantial & Weak to moderate & Moderate & Moderate \\
$1/10$ to $1/30$ & Strong & Moderate to strong & Substantial & Strong \\
$1/30$ to $1/100$ & Very strong & Strong & Strong & Very strong\\
$1/100$ to $1/300$ & Decisive & Very strong & Very strong & Extreme\\
$<1/300$ & & & Decisive & \\
\hline  
\end{tabular}
\end{table*}
\label{tab:bfs}
For example, according to table 1, a Bayes factor $BF_{01}=1/5$ can be interpreted as moderate evidence against the null hypothesis $H_0$ relative to the alternative hypothesis $H_1$ when the scale of Held \& Ott \citeyear{Held2016} is used. Using the relationship $BF_{01}=1/BF_{10}$, one obtains $BF_{10}=5$, and the Bayes factor $BF_{10}=5$ indicates a necessary change in beliefs towards $H_1$. Again, this highlights the appealing property of Bayes factors that it is always possible to express the evidence \textit{for} a research hypothesis, in contrast to $p$-values which can only \textit{reject} a null hypothesis. Bayes factors can be used to \textit{confirm} a research hypothesis under consideration.

However, an often lamented problem with the Bayes factor as discussed by Kamary et al. \citeyear{Kamary2014} and Robert \citeyear{Robert2016} is its dependence on the prior distributions which are assigned to the model parameters. While this criticism is valid, modern software like JASP \shortcite{Jasp2019,VanDoorn2019} enables researchers to use so-called robustness analyses which show how the resulting Bayes factor changes when the prior distribution is varied. Using such robustness analyses, it is possible to check if the resulting Bayes factor is highly unstable or if the prior assumptions have little effect on the magnitude of the obtained Bayes factor.

To apply the Bayes factor in the running example, there are two options: First, it is possible to use a closed-form expression to calculate the Bayes factor based on the observed data $x$. For details, see Rouder et al. \citeyear{Rouder2009}. The resulting Bayes factor is $BF_{01}=0.6870$, which according to table 1 does not signal evidence \textit{against} $H_0$. The Bayes factor for $H_1$, given as $BF_{10}=1.4556$ shows that the evidence against $H_0$ is also bare worth mentioning. In summary, the Bayes factor indicates that data is indecisive in the running example.

A second option to obtain the Bayes factor numerically is the Savage-Dickey density ratio method \cite{Dickey1970,Verdinelli1995}. A concise introduction for the Savage-Dickey density ratio for psychologists is given by Wagenmakers et al. \citeyear{Wagenmakers2010}. The Savage-dickey density ratio states that the Bayes factor $BF_{01}$ can be obtained numerically as the ratio of the posterior density's value at $\delta_0$ and the prior density's value at $\delta_0$:
\begin{align}
    BF_{01}=\frac{p(\delta_0|x,H_1)}{p(\delta_0,H_1)}    
\end{align}
In general, this relationship holds for any parameter $\theta$ of interest, and in the running example the parameter of interest is the effect size $\delta$. In the running example one obtains for $\delta_0=0$ the values $p(\delta_0|x,H_1)=0.2171$ and $p(delta_0|H_1)=0.3183$ so that $BF_{01}=0.2171/0.3183=0.6821$, which is very close to the analytically obtained Bayes factor of $BF_{01}=0.6870$.

\begin{figure}[!h]
\centering
\includegraphics[width=0.5\textwidth]{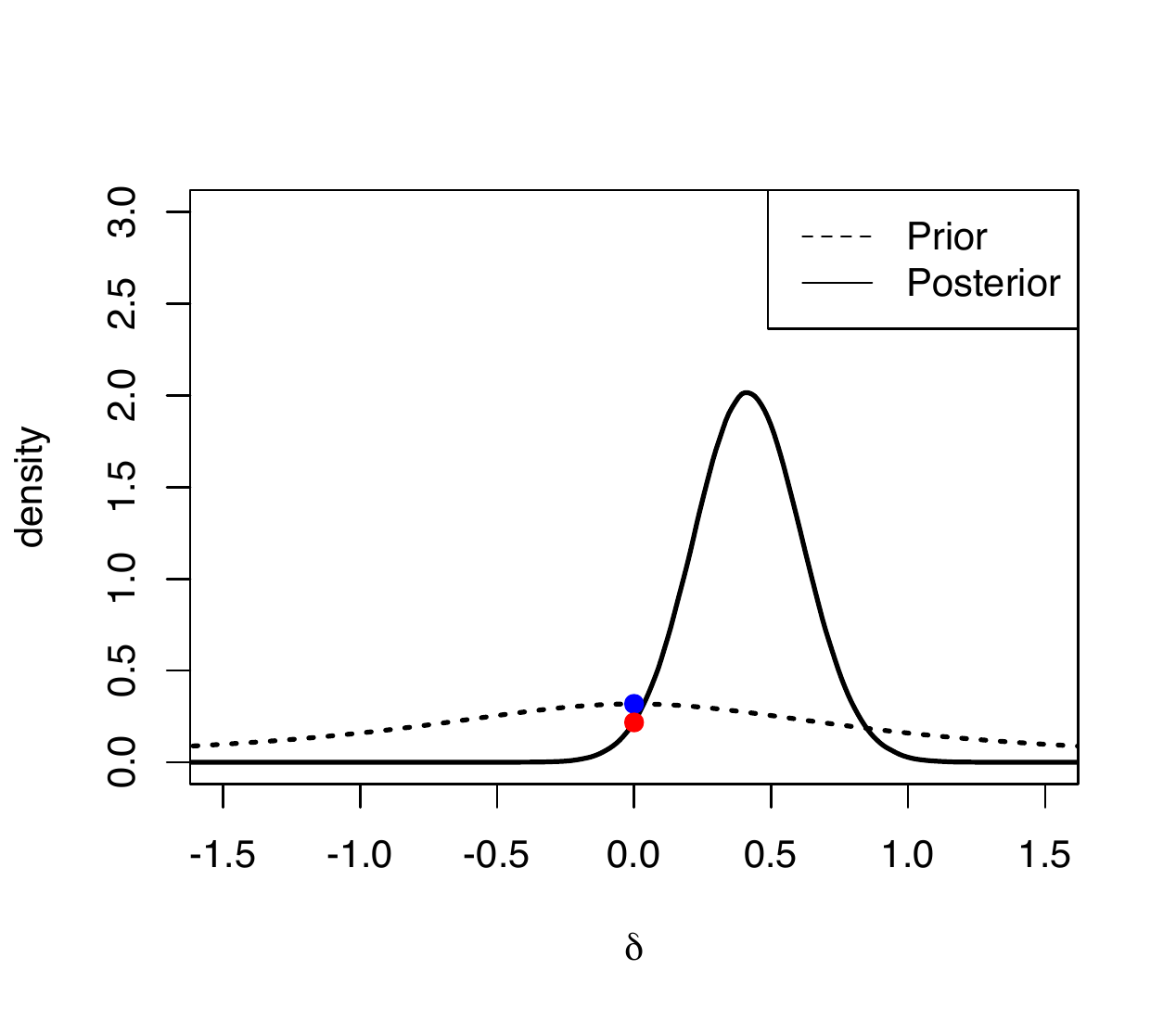}
\caption{Prior-posterior plot for the effect size of the two-sample Bayesian t-test in the running example. Dashed line shows the prior distribution $p(\delta)=C(0,1)$, which is a wide Cauchy prior distribution. Solid line shows the posterior distribution $p(\delta|x)$ after observing the data $x$ in both groups. The \textit{Savage-Dickey density ratio} is visualised by the red and blue points: The Bayes factor $BF_{01}$ is the ratio of the posterior density's height (red point) and the prior density's height (blue point) at the null value $\delta_0=0$.}
\label{fig:1}
\end{figure}

In summary, the Bayes factor is an appealing option to test hypotheses in the Bayesian paradigm. It can state evidence both for the null and the alternative hypothesis, robustness analyses prevent the results to be highly dependent on the prior assumptions made, and calculations for most standard models used in psychological research are either straightforward or obtained via numerical methods like the Savage-Dickey density ratio. Notice that it is also possible to obtain posterior probabilities of each hypothesis explicitly via equation (\ref{eq:bf}). These aspects contribute to the fact that the Bayes factor is already widely used among mathematical psychologists \shortcite{Hoijtink2019} and has influenced scientific practice in psychology \shortcite{Dienes2016,Schoenbrodt2017}. However, the dependence on the parameter prior distributions and computational difficulties to obtain the marginal likelihoods are disadvantages of the Bayes factor \cite{Tendeiro2019}.

\subsection{The region of practical equivalence (ROPE)}
The Bayes factor was designed to test a precise hypothesis $H_0:\theta=\theta_0$, where $\theta_0$ is the value of interest. The increasing reproducibility problems of frequentist hypothesis tests and $p$-values in psychology led Cumming \citeyear{Cumming2014} to propose a shift from hypothesis testing to estimation under uncertainty. Instead of testing a precise hypothesis $H_0:\theta=\theta_0$ via a significance test which employs p-values, he advocated calculating interval estimates like confidence intervals instead. The plausibility of a hypothesis $H_0:\theta=\theta_0$ is subsequently judged by the uncertainty of the interval estimate: If the null hypothesis parameter $\theta_0$ is included in the confidence interval, $H_0$ cannot be rejected. If $\theta_0$ is located outside of the confidence interval, one can reject $H_0$. Cumming \citeyear{Cumming2014} additionally included in his proposal a focus on \textit{``estimation based on effect sizes''} \cite[p.~7]{Cumming2014} instead of $p$-values and hypothesis tests, see also Greenland et al. \citeyear{Greenland2016}.

This shift was proposed originally from frequentist hypothesis testing to frequentist estimation and dubbed the \textit{'New Statistics'}. Conceptually, the idea is appealing, and such a process is indeed observed across multiple scientific domains by now \cite{Wasserstein2019}. Also, such a shift towards an estimation-oriented perspective embraces uncertainty instead of the dichotomous separation of research results via $p$-values into significant and non-significant findings. This is closely related to one of the six principles for properly interpreting $p$-values which were stated in the 2016 statement of the American Statistical Association to the reproducibility crisis. It underlined that a $p$-value \textit{``does not measure the size of an effect or the importance of a result.''} \cite[p.~132]{wasserstein2016}. Instead, the original proposal of Cumming \citeyear{Cumming2014} focusses on effect sizes and estimation instead of hypothesis testing.

Kruschke \& Liddell \citeyear{Kruschke2018} proposed a similar shift towards estimation under uncertainty from a Bayesian perspective. They argued that as frequentist interval estimates like confidence intervals are obtained by inverting a frequentist hypothesis test \cite{Casella2002a}, the problems inherent to significance testing via $p$-values are inherited by confidence intervals. As a consequence, confidence intervals as quantities for estimation are \textit{``highly sensitive to the stopping and testing intentions.''} \cite[p.~184]{Kruschke2018}, while Bayesian interval estimates due to the accordance of Bayesian statistics with the likelihood principle are not. To avoid these problems, Kruschke \& Liddell \citeyear{Kruschke2018} argued that the shift originally proposed by Cumming \citeyear{Cumming2014} can be achieved easier by Bayesian data analysis.


Next to preferring Bayesian instead of frequentist interval estimates for such a shift, Kruschke \& Liddell \citeyear{Kruschke2018b} argued that \textit{precise} hypothesis testing as achieved via the Bayes factor is not suitable in a variety of realistic research settings. Examples include exploratory research, measurements which include a substantial amount of error, and in general complex phenomena in which a precise hypothesis $H_0:\theta=\theta_0$ at best can be interpreted as an approximation to reality \shortcite{Berger1987,Good1988,Berger1997}. For example, Rouder et al. \citeyear{Rouder2009} argued:
\begin{quote}
    ``It is reasonable to ask whether hypothesis testing is always necessary. In many ways, hypothesis testing has been employed in experimental psychology too often and too hastily (...). To observe structure, it is often sufficient to plot estimates of appropriate quantities along with measures of estimation error (Rouder \& Morey, 2005). As a rule of thumb, hypothesis testing should be reserved for those cases in which the researcher will entertain the null as theoretically interesting and plausible, at least \textit{approximately}.''\newline
    \shortcite[p. 235]{Rouder2009}
\end{quote}
Notice that in the frequentist approach, \textit{precise} hypothesis testing can be interpreted as \textit{null} hypothesis significance testing, and the null hypothesis $H_0:\theta=\theta_0$ then \textit{precisely} states that $\theta=\theta_0$. In the Bayesian paradigm, the same idea is sometimes called \textit{sharp} hypothesis testing \cite{Pereira2008}. Conceptually, both names identify the same practice. However, in psychological research the assumption of a precise effect is highly questionable as Berger et al. \citeyear{Berger1997} noted:
\begin{quote}
        ``The decision whether or not to formulate an inference problem as one of testing a precise null hypothesis centers on assessing the plausibility of such an hypothesis. Sometimes this is easy, as in testing for the presence of extrasensory perception, or testing that a proposed law of physics holds. Often it is less clear. In medical testing scenarios, for instance, it is often argued that any treatment will have some effect, even if only a very small effect, and so exact equality of effects (between, say, a treatment and a placebo) will never occur.''\newline
        \cite[p.~145]{BergerBrownWolpert1994}
\end{quote}
Based on these ideas, Kruschke \& Liddell \citeyear{Kruschke2018b} considered \textit{approximate} hypothesis testing instead of \textit{precise} hypothesis testing. To facilitate the shift to an estimation-oriented perspective in the veins of Cumming \citeyear{Cumming2014}, they proposed the \textit{region of practical equivalence (ROPE)}. The ROPE unites a concept which appears under different names in various scientific areas,  \textit{`including ``interval of clinical equivalence'', ``range of equivalence'', ``equivalence interval'', ``indifference zone''}, \textit{``smallest effect size of interest,''} \textit{ and ``good-enough belt'' ...'} \cite[p.~185]{Kruschke2018}, where these terms come from a wide spectrum of scientific contexts, see Carlin \& Louis \citeyear{Carlin2009}, Freedman et al. \citeyear{Freedman1983}, Hobbs \& Carlin \citeyear{Hobbs2007}, Lakens \citeyear{Lakens2014,Lakens2017} and Schuirmann \citeyear{Schuirmann1987}. The general idea of all these concepts is to establish a region of practical equivalence around the null value $\theta_0$ of the null hypothesis $H_0:\theta=\theta_0$, which expresses \textit{``the range of parameter values that are equivalent to the null value for current practical purposes.''} \cite[p.~185]{Kruschke2018}. To test hypotheses via the ROPE and a Bayesian interval estimate, the highest posterior density (HPD) interval, Kruschke \citeyear{Kruschke2018a} proposed the following decision rule:
\begin{itemize}
    \item{Reject the null value $\theta_0$ specified by $H_0:\theta=\theta_0$, if the 95\% HPD falls \textit{entirely} outside the ROPE.}
    \item{Accept the null value, if the 95\% HPD falls \textit{entirely} inside the ROPE.}    
\end{itemize}
If the 95\% HPD falls \textit{entirely} outside the ROPE, the parameter value is not inside the ROPE with more than 95\% posterior probability, and as a consequence not \textit{practically equivalent} to the null value $\theta_0$. A rejection of the null value $\theta_0$ and the corresponding hypothesis then seems reasonable.

If the 95\% HPD falls \textit{entirely} inside the ROPE, the parameter value is located inside the ROPE with at least 95\% posterior probability. As a consequence, it is \textit{practically equivalent} to the null value $\theta_0$ and it is reasonable to accept the null hypothesis $H_0:\theta=\theta_0$.

Formally, the decision rule can be derived as a Bayes rule \cite{Held2014} for a specific choice of a loss function. Also, the decision based on the HPD and ROPE approach to Bayesian hypothesis testing is asymptotically consistent. For details, see the supplement file in Kruschke \citeyear{Kruschke2018a}.\footnote{Formally, Kruschke \citeyear{Kruschke2018a} avoided the expression hypothesis testing when using the ROPE and HPD approach, as the goal is estimation under uncertainty and not precise hypothesis testing. We agree, and when we use the phrase hypothesis testing here in connection with the ROPE, we always mean \textit{approximate} hypothesis testing, which questions the use of precise hypothesis testing in a variety of research settings and emphasises estimation over testing.} Also, the decision rule is described in more detail in Kruschke et al. \shortcite{Kruschke2012,Kruschke2015a} and Kruschke \& Liddell \cite{Kruschke2018b}.



At first, using the ROPE in combination with a Bayesian posterior distribution seems appealing. However, there are two limitations which have prevented more widespread use in psychological research by now:
\begin{enumerate}[label=(\arabic*)]
    \item{Hypothesis testing based on the ROPE and HPD interval is only possible in situations where scientific standards of practically equivalent parameter values exist and are widely accepted by the research community.}
    \item{The ROPE only measures the concentration of the posterior's probability mass, but fails to measure evidence once the entire HPD is located inside or outside the ROPE.}
    \end{enumerate}
    
The first limitation is less severe, as for a variety of quantities there exist widely accepted standards on how to interpret different magnitudes. Examples are given by effect sizes, which are categorised in psychology, for example, according to Cohen \citeyear{cohen_statistical_1988}. A widely accepted ROPE $R$ around a null hypothesis $H_0:\delta=0$ for the effect size $\delta$ is given as $R=[-0.1,0.1]$ \cite{Makowski2019,Kelter2020BayesianPosteriorIndices}. The boundaries $\delta=-0.1$ and $\delta=0.1$ are precisely half of the magnitude necessary for at least a small effect according to Cohen \citeyear{cohen_statistical_1988}. Similar proposals for default ROPEs have been made for logistic and linear regression models. There, a standard ROPE is $|\beta|\leq 0.05$ for regression coefficients, see \cite[p.~277]{Kruschke2018a}. These default values are justified both by mathematical theory and official guidelines. The default ROPEs above are inspired by the recommendations of the U.S. Food and Drug Administration's Center for Drug Evaluation and Research and Center for Biologics Evaluation and Research \citeyear{USFoodAndDrugAdministrationBioequivalence2001}. Also, they are based on the official guidelines of the Center for Veterinary Research \citeyear{UnitedStatesFoodandDrugAdministrationCenterforVeterinaryMedicine2016} and the recommendations of the Center for Biologics Evaluation and Research \citeyear{UnitedStatesFoodandDrugAdministrationCenterforDrugEvaluationandResearchandCenterforBiologicsEvaluationandResearch}.

Additionally, the default ROPEs behave similarly to the Bayes factor in practice as shown by Makowski et al. \citeyear{Makowski2019} and Kelter \citeyear{Kelter2020BayesianPosteriorIndices}.

In summary, there exist theoretically justified default values for determining the boundaries of the ROPE for widely used parameters in psychological research \shortcite{Westlake1976,Kirkwood1981, Lakens2017,Kruschke2018a,Makowski2019,Kelter2020BayesianPosteriorIndices} making the ROPE and HPD approach a considerable alternative method for Bayesian hypothesis testing.

\begin{figure}[!h]
\centering
\includegraphics[width=0.5\textwidth]{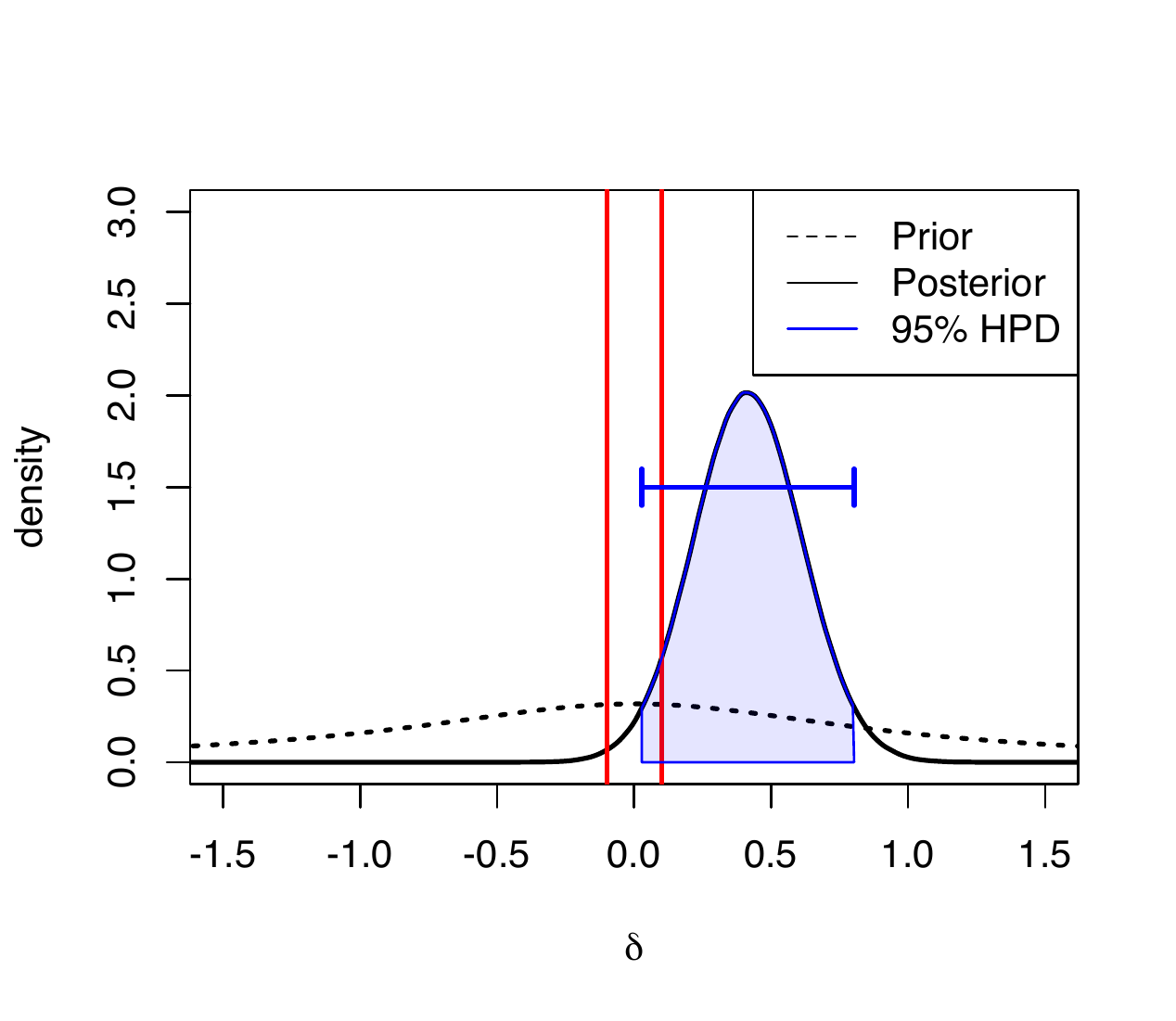}
\caption{The ROPE and HPD approach for the effect size $\delta$ of the two-sample Bayesian t-test in the running example. The vertical red lines show the boundaries of the ROPE $R=[-0.1,0.1]$, and the blue interval is the 95\% HPD interval. The blue shaded area corresponds to the posterior probability mass inside the 95\% HPD, and from the figure it is clear that only a tiny amount of probability mass (3.16\%) is located inside the ROPE $R$.}
\label{fig:rope}
\end{figure}

The second problem is more severe, as the ROPE cannot separate between two situations in which the HPD has concentrated entirely inside the ROPE. However, the concentration around the null value $\theta_0$ may be much more apparent in one case, while in the other the HPD boundaries coincide with the ROPE boundaries, stating much less evidence for the null hypothesis. The same situation holds when the HPD is located entirely outside the ROPE.

In the running example, figure \ref{fig:rope} shows the situation using the default ROPE $R=[-0.1,0.1]$ around the effect size null value $\delta_0=0$. The 95\% HPD interval ranges from $\delta=0.03$ to $\delta=0.80$, and therefore is not located entirely outside the ROPE $R$. As a consequence, the null hypothesis cannot be rejected. Notice that when considering the amount of posterior probability mass located inside the ROPE $R$, only $3.16\%$ are located inside the ROPE $R$. Choosing a $90\%$ HPD interval instead in combination with the ROPE $R$ would lead to the rejection of $H_0:\delta=0$.

\subsection{The MAP-based $p$-value}
The MAP-based $p$-value was proposed by Mills \citeyear{Mills2017}, and is the ratio of the posterior densities null value $p(\theta_0)$ and the posterior densities maximum a posteriori (MAP) value:
\begin{align*}
    p_{\text{MAP}}=\frac{p(\theta_0|x)}{\arg \max\limits_{\theta \in \Theta} p(\theta|x)}
\end{align*}
The idea is based on the likelihood ratio used in the Neyman-Pearson theory \cite{Neyman1933,Casella2002a}. However, here the denominator is not maximising the likelihood over the alternative hypothesis $H_1$ like in a traditional Neyman-Pearson test, but instead the posterior density $p(\theta|x)$ is maximised.

\begin{figure}[!h]
\centering
\includegraphics[width=0.5\textwidth]{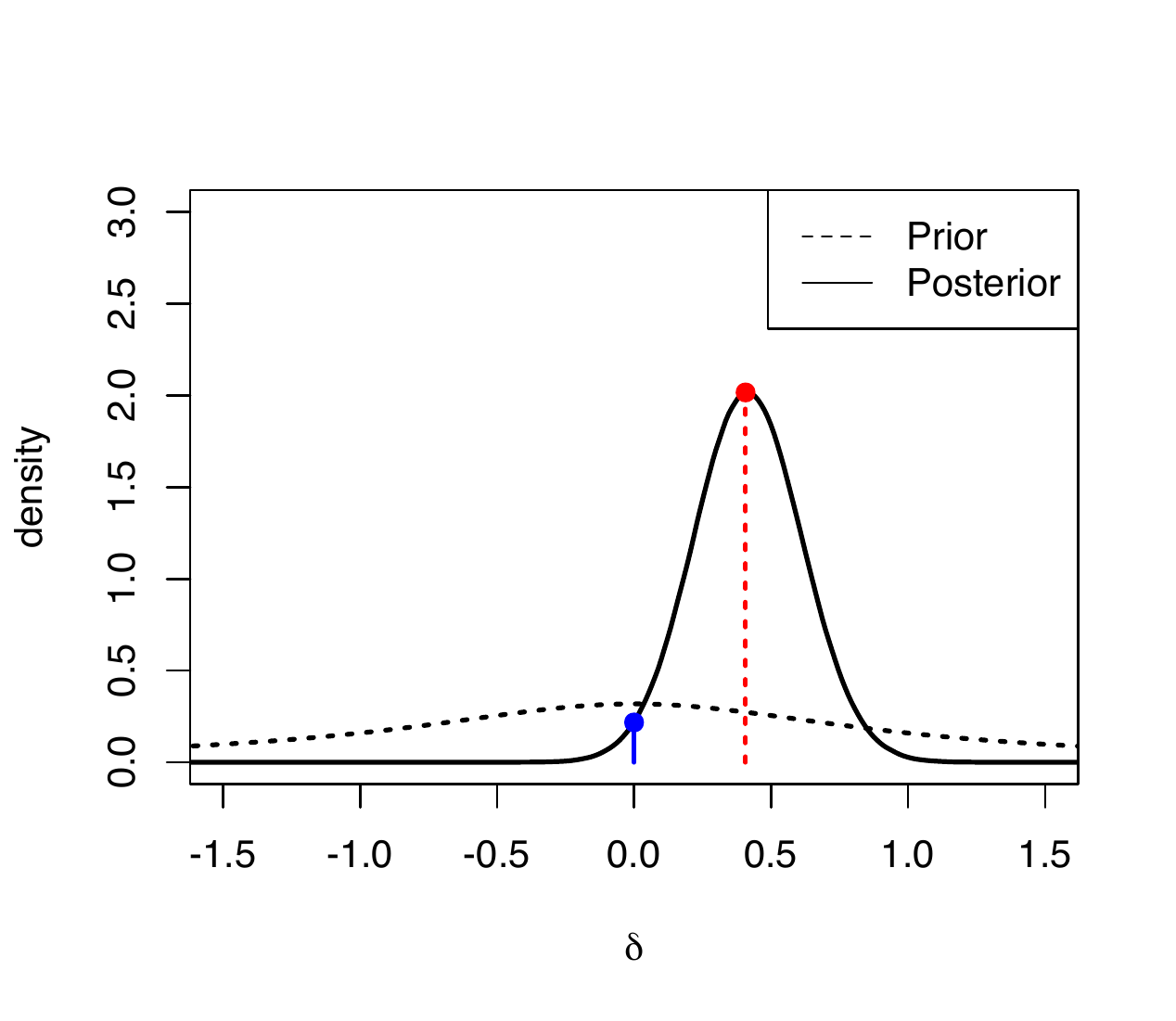}
\caption{The MAP-based $p$-value for the effect size $\delta$ of the two-sample Bayesian t-test in the running example. The solid blue line shows the posterior distribution's value $p(\delta=0|x)$ at the null value $\delta=0$. The dashed red line shows the posterior distribution's MAP value $p(\delta=0.41|x)$ at the MAP value $\delta_{\text{MAP}}=0.41$. The MAP-based $p$-value is given as the ratio between these two values, and is significant if $p_{\text{MAP}}<0.05$.}
\label{fig:mapBasedPValue}
\end{figure}

In the running example, $\theta$ is the effect size $\delta$, the null hypothesis value is $\delta_0=0$, and the MAP-based $p$-value becomes
\begin{align*}
    p_{\text{MAP}}=\frac{p(\delta=0|x)}{\arg \max\limits_{\delta \in \mathbb{R}} p(\delta|x)} = \frac{p(\delta=0|x)}{p(\delta=0.41|x)} \approx 0.1076
\end{align*}
Figure \ref{fig:mapBasedPValue} visualises the MAP-based $p$-value: The solid blue line is the posterior distribution's value $p(\delta=0|x)$ at the null value $\delta=0$. The dashed red line is the posterior distribution's MAP value $p(\delta=0.41|x)$ at the MAP value $\delta_{\text{MAP}}=0.41$. The MAP-based $p$-value is the ratio between these two values, and is significant if $p_{\text{MAP}}<0.05$. 

As $p_{\text{MAP}}> 0.05$, the null hypothesis $H_0:\delta = 0$ is not rejected, similar to the idea behind traditional $p$-values.

Notice that the MAP-based $p$-value can only be used to reject a null hypothesis $H_0$ similar to traditional $p$-values. It is not possible to confirm a hypothesis. Also, there is by now no associated loss function which identifies the MAP-based $p$-value as a Bayes rule which minimises the expected loss under the selected loss function. Also, while decision thresholds (or test levels $\alpha$) of frequentist $p$-values are theoretically justified to bound the type I error rate while simultaneously minimising the type II error rate -- leading to the concept of \textit{uniformly most powerful} (UMP) level $\alpha$ tests \cite{Neyman1938,Casella2002a} -- there is no analogue justification for choosing a specific threshold for MAP-based $p$-values.

\subsection{The probability of direction}
The probability of direction (PD) is formally defined as the proportion of the posterior distribution that is of the median's sign. Although expressed differently, this index is strongly correlated to the frequentist p-value \shortcite{Makowski2019}.

\begin{figure}[!h]
\centering
\includegraphics[width=0.5\textwidth]{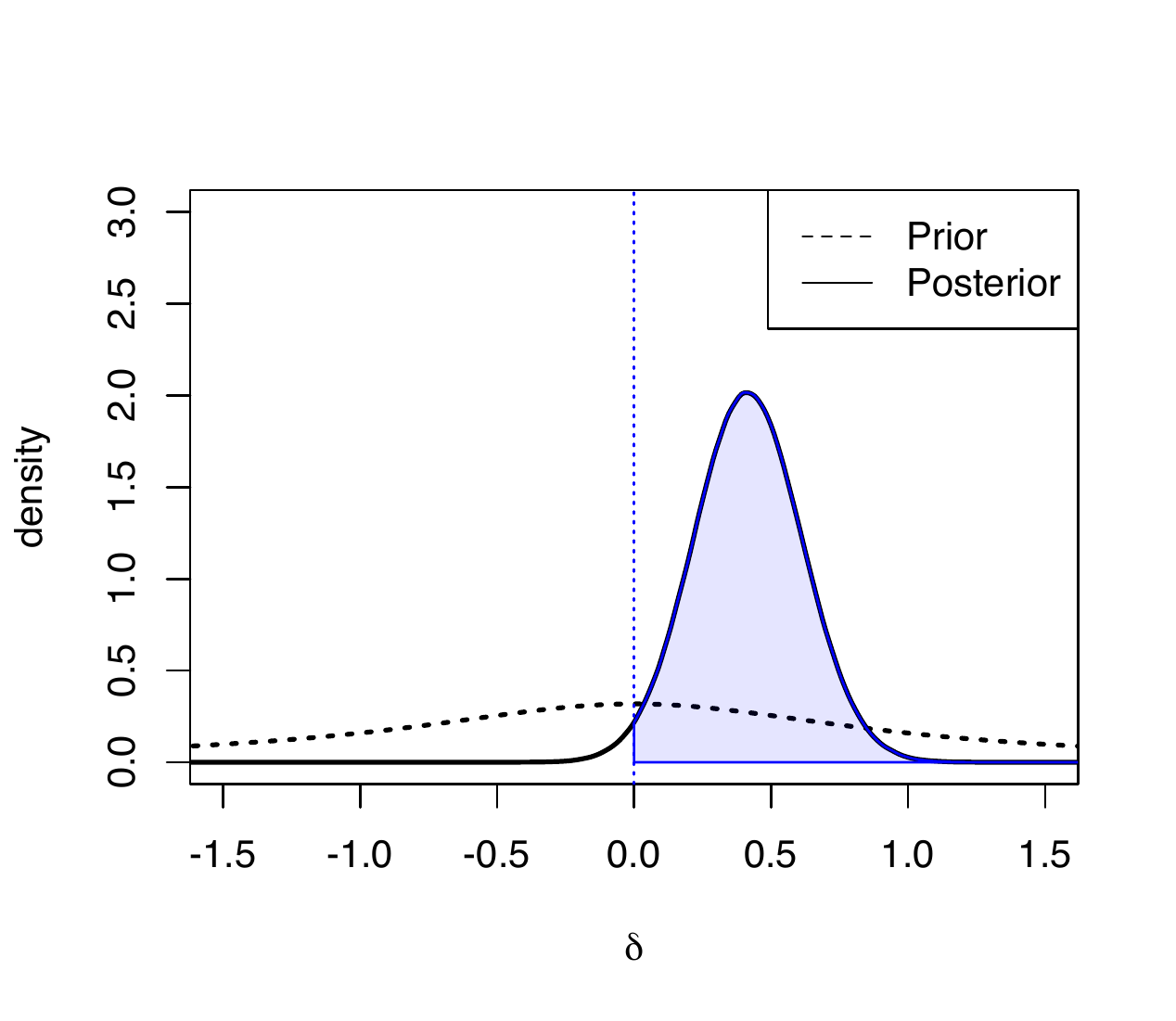}
\caption{The probability of direction (PD) for the effect size $\delta$ of the two-sample Bayesian t-test in the running example. The blue dotted line separates posterior parameter values with a positive and negative sign from each other. The shaded blue area is the proportion of the posterior which is of the median's sign.}
\label{fig:pd}
\end{figure}

In the running example, the posterior median of the effect size $\delta$ is given as $\delta_{\text{MED}}=0.41$, which has a positive sign. Therefore, the PD is simply the proportion of posterior probability mass which is of a positive sign:
\begin{align*}
    PD = \int_{\mathbb{R}_{+}} p(\delta|x)d\delta \approx 0.9827
\end{align*}
Figure \ref{fig:pd} visualises the PD in the running example: The blue dotted line separates positive from negative posterior parameter values. The blue shaded area under the posterior distribution is the proportion of posterior probability mass which is of the median's sign, which equals the PD.
Based on the PD one can test the null hypothesis $H_0:\delta=0$ by requiring a specified amount of posterior probability mass to be strictly positive or negative. For example, if $PD=1$, all posterior parameter values are positive (or negative) and rejection of $H_0$ seems reasonable. In the running example, more than 95\% of the posterior indicate that there is a positive effect, and if one uses this threshold for deciding between $H_0$ and $H_1$, $H_0$ would be rejected.

However, similar to the MAP-based $p$-value there exists no loss function which identifies the PD as a Bayes rule which minimises the expected loss under the selected loss function. Also, there are no theoretically justified thresholds which makes the decision based on the PD arbitrary. Note however that this criticism is also valid for Bayes factors and the ROPE and HPD approach, as the selection of a specific ROPE and HPD or the translation of different sizes of Bayes factors into evidence is arbitrary, too \cite{Tendeiro2019}.

\subsection{The Full Bayesian Significance Test (FBST) and the $e$-value}
The Full Bayesian Significance Test (FBST) was developed by Pereira and Stern \citeyear{Pereira1999} as a Bayesian alternative to frequentist significance tests employing the $p$-value. It was designed to test a \textit{precise} hypothesis.\footnote{Formally, Pereira et al. \citeyear{Pereira2008} referred to a \textit{sharp} hypothesis which is defined as any submanifold of the parameter space of interest. This includes for example point hypotheses like $H_0:\delta=0$ or in general hypotheses which consist of a set of lower dimension than the parameter space.} The FBST can be used with any standard parametric model, where $\theta \in \Theta \subseteq \mathbb{R}^p$ is a (possibly vector-valued) parameter of interest, $p(x|\theta)$ is the likelihood and $p(\theta)$ is the prior distribution.

A precise hypothesis $H_0$ makes the statement that the parameter $\theta$ lies in the corresponding null set $\Theta_{H_0}$. For point null hypotheses like $H_0:\theta=\theta_0$ the null set simply is given as $\Theta_{H_0} = \theta_0$. The idea of the FBST is to use the $e$-value, which quantifies the Bayesian evidence against $H_0$ as a Bayesian replacement of the traditional $p$-value. To construct the $e$-value, Stern \& Pereira \citeyear{Stern2020} defined the posterior \textit{surprise function} $s(\theta)$:
\begin{align}
    s(\theta):=\frac{p(\theta|x)}{r(\theta)} 
\end{align}
The surprise function normalises the posterior distribution $p(\theta|x)$ by a reference function $r(\theta)$. Possible choices include a flat reference function $r(\theta)=1$ or any prior distribution $p(\theta)$ for the parameter $\theta$, that is: $r(\theta)=p(\theta)$. In the first case, the surprise function becomes the posterior distribution $p(\theta|x)$, and in the second case parameter values $\theta$ with a surprise function $s(\theta)\geq 1$ indicate that they have been corroborated by the data $x$, while parameter values $\theta$ with a surprise function $s(\theta)<1$ indicate that they have become less probable a posteriori. The supremum $s^{*}$ is then defined as the supremum of the surprise function $s(\theta)$ over the null set $\Theta_{H_0}$ which belongs to the hypothesis $H_0$ :
\begin{align*}
    s^{*}:=s(\theta^{*})=\sup\limits_{\theta \in \Theta_{H_0}}s(\theta)
\end{align*}
Pereira \& Stern \citeyear{Pereira1999} then defined the tangential set $\overline{T}(\nu)$ to the hypothesis $H_0$ as
\begin{align*}
    \overline{T}(\nu):=\Theta \setminus T(\nu)
\end{align*}
where
\begin{align*}
    T(\nu):=\{\theta \in \Theta|s(\theta)\leq \nu \}
\end{align*}
$T(s^{*})$ includes all parameter values $\theta$ which are smaller or equal to the supremum $s^{*}$ of the surprise function under the null set, and $\overline{T}(s^{*})$ includes all parameter values $\theta$ which are larger than the supremum $s^{*}$ of the surprise function under the null set.

\begin{figure*}[!h]
\centering
\includegraphics[width=\textwidth]{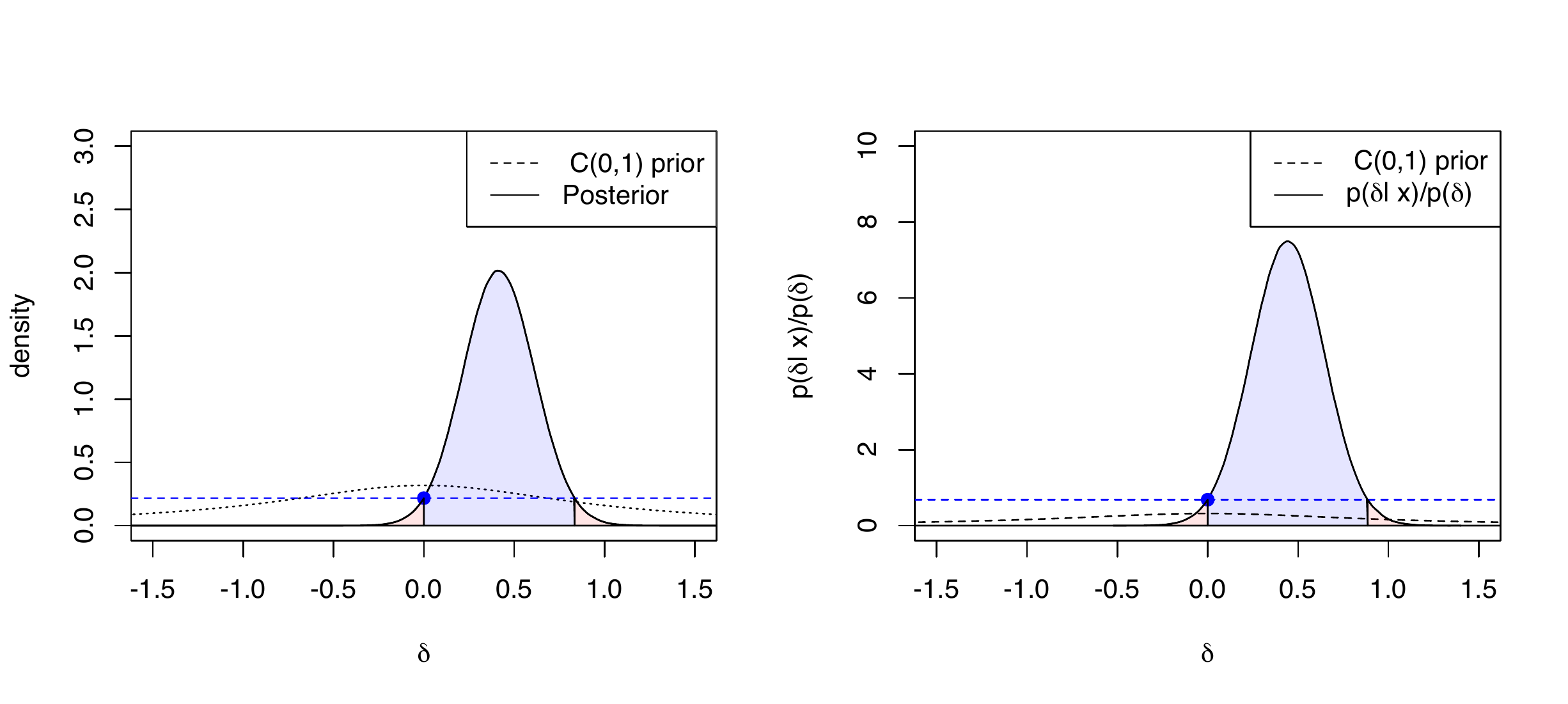}
\caption{The FBST and the $e$-value $\overline{\text{ev}}(H_0)$ against $H_0$ when using a flat reference function $r(\delta)=1$ (left) and a wide Cauchy reference function $r(\delta)=C(0,1)$ (right) for the Bayesian two-sample t-test; In both cases, a wide $C(0,1)$ prior was assigned to the effect size $\delta$; the blue shaded area is the integral over the tangential set $\overline{T}(0)$ against $H_0:\delta =0$, which is the $e$-value $\overline{\text{ev}}$ against $H_0$; the red area is the integral over $T(0)$, which is the $e$-value ev$(H)$ in favour of $H_0:\delta=0$}
\label{fig:fbst}
\end{figure*}

The last step towards the $e$-value is then to define the \textit{cumulative surprise function} $W(\nu)$
\begin{align}
    W(\nu):=\int_{T(\nu)}p(\theta|x)d\theta
\end{align}
and setting $\nu=s^{*}$, $W(s^{*})$ is simply the integral of the posterior distribution $p(\theta|x)$ over $T(s^{*})$. The Bayesian $e$-value, which measures the evidence \textit{against} the null hypothesis $H_0$, is then defined as
\begin{align}
    \overline{\text{ev}}(H_0):=\overline{W}(s^*)
\end{align}
where $\overline{W}(\nu):=1-W(\nu)$. The left plot in figure \ref{fig:fbst} visualises the FBST and the $e$-value $\overline{\text{ev}}(H_0)$ in the running example. As previously, the dashed line shows the $C(0,1)$ Cauchy prior on the effect size $\delta$, and the solid line is the resulting posterior distribution $p(\delta|x)$.

A flat reference function $r(\delta)=1$ was selected, so that the posterior surprise function $s(\delta)$ becomes the posterior distribution $p(\delta|x)$.

The supremum of the surprise function $s(\delta)$ over the null set $\Theta_{H_0}=\{\delta_0\}$ becomes $s^{*}=0$, because $H_0:\delta=\delta_0$ with $\delta_0=0$. Therefore, the null set $\Theta_{H_0}=\{\delta_0\}$ includes only the single value $\delta_0=0$, which then also is the supremum. The blue point visualises the value $s^{*}=s(0)\approx 0.2171$.

The tangential set $\overline{T}(s^{*})$ is then simply the set of parameter values which satisfy $s(\delta) > s(0)$ (that is, $s(\delta) > 0.2171)$), and the horizontal blue dashed line shows the boundary between parameter values $\delta$ with $s(\delta) > s(0)$ (that is, $s(\delta) > 0.2171)$) and $s(\delta) \leq s(0)$ (that is, $s(\delta) \leq 0.2171)$).

The value of the cumulative surprise function $W(s^{*})=W(0.2171):=\int_{T(0.2171)}p(\theta|x)d\theta$ is then visualised as the red shaded area under the posterior, which is the integral of the posterior over all parameter values $\delta$ which are smaller or equal to $s^{*}=s(0)=0.2171$.

The $e$-value $\overline{\text{ev}}(H_0)$ against $H_0$ is then given as $\overline{W}(s^*)=1-W(s^{*})=1-W(s(0))$, which is the integral of the posterior over all parameter values $\delta$ which are larger than $s^{*}=s(0)=0.2171$, visualised as the blue shaded area under the posterior. In the running example, $\overline{\text{ev}}(H_0)=0.9659$, which means that $96.59\%$ of the posterior distribution's parameter values attain a larger posterior density value than the null value $\delta_0=0$. As a consequence, the null hypothesis $H_0$ traverses (or is located) in a low posterior density region, and the Bayesian evidence against $H_0:\delta=0$ is substantial. If one would use a threshold of $95\%$, one could reject the null hypothesis $H_0$ based on the $e$-value in the running example.

The right plot in figure \ref{fig:fbst} shows the surprise function $s(\delta)$ when a wide Cauchy prior $p(\delta)=C(0,1)$ is selected as reference function $r(\delta)$. The resulting $e$-value against $H_0$ in this case is $\overline{\text{ev}}(H_0)=0.9743$.

Notice that formally, the $e$-value ev$(H_0)$ \textit{supporting} $H_0$ is obtained as ev$(H_0):=1-\overline{\text{ev}}(H_0)$. However, the Bayesian evidence for $H_0$, the $e$-value $\text{ev}(H_0)$ is not evidence against $H_1$, because $H_1$ is not even a sharp hypothesis.\footnote{See Definition 2.2 in Pereira et al. \citeyear{Pereira2008}.}.

Also, it is not possible to use the $e$-value ev$(H_0)$ to \textit{confirm} $H_0$ in the sense of a Bayes factor or ROPE. For details see Kelter \citeyear{Kelter2020BayesianPosteriorIndices}. Nevertheless, one can use ev$(H_0)$ to reject $H_0$ if ev$(H_0)$ is sufficiently small, and there are asymptotic arguments based on the distribution of ev$(H_0)$ which make it possible to obtain critical values to reject a hypothesis $H_0$ similar to $p$-values in frequentist significance tests \shortcite{Stern2020}. Therefore, in practice, one could alternatively use these critical values of the distribution of ev$(H_0)$ instead of using an (arbitrary) threshold like $95\%$ for ev$(H_0)$ as used above.

In summary, the FBST and the $e$-value were designed to mimic a frequentist significance test of a \textit{precise} hypothesis. The $e$-value $\overline{\text{ev}}(H)$ can be interpreted as a direct replacement of the frequentist $p$-value and can only be used to reject a null hypothesis $H_0$ of interest. The confirmation of a research hypothesis via ev$(H)$ is not possible \cite{Kelter2020BayesianPosteriorIndices, Stern2020}.

However, the FBST has appealing properties: First, it is ``Bayesian'' in the sense that its inference procedure can be derived by minimising an appropriate loss function \shortcite{Madruga2001,Madruga2003}. Second, the $e$-value is explicitly invariant under suitable transformations on the coordinate system of the parameter space \cite{Pereira2008}. Third, the FBST works in the original parameter space and encounters no problem when nuisance parameters are present \cite[Section 3.1]{Stern2020}. In contrast, the computation of the marginal likelihoods for obtaining the Bayes factor can quickly become difficult in these situations. Also, the FBST and $e$-value have been widely applied in a variety of scientific domains over the last two decades \cite{Stern2020}. Interestingly, the FBST has to our best knowledge not been applied in psychology so far.

\section{Comparison}
Table \ref{tab:comparison} shows a comparison of the various available posterior indices for hypothesis testing in the Bayesian paradigm.
\begin{table*}[bt]
\caption{Comparison of different Bayesian posterior indices for hypothesis testing}
\begin{tabular}{p{1.5cm}p{4cm}p{2.3cm}p{4cm}p{3.5cm}}
\toprule
Posterior index & Typical research question & Can be used for... & Benefits & Limitations\\
\midrule
Bayes factor & Which change in relative beliefs about both hypotheses $H_0$ and $H_1$ is required after observing the data $x$? & $\checkmark $ Confirmation\newline
    $\checkmark $ Rejection\newline
    of a hypothesis & $\checkmark $ Can be derived as an explicit Bayes' rule\newline
$\checkmark $ Easy to interpret and well-established & \xmark $ $ Scales for interpretation are arbitrary\newline
    \xmark $ $ Computation of marginal likelihoods can be difficult in presence of nuisance parameters\\
\hline
ROPE & Is the parameter $\theta$ practically equivalent to values inside the ROPE $R$? & $\checkmark $ Confirmation\newline
    $\checkmark $ Rejection\newline
    of a hypothesis & $\checkmark $ Can be derived as an explicit Bayes' rule\newline
 $\checkmark $ Easy to interpret\newline
 $\checkmark $ Allows to establish bio-equivalence which is more realistic than testing for precise effects & \xmark $ $ Selection of default ROPEs is not straightforward for all parameters\newline
 \xmark $ $ Cannot measure evidence once the HPD has concentrated entirely inside or outside the ROPE\\
 \hline
MAP-based $p$-value & How probable is null value $\theta_0$ compared to the MAP value $\theta_{\text{MAP}}$? & \xmark $ $ Confirmation\newline
    $\checkmark $ Rejection\newline
    of a hypothesis & $\checkmark $ Allows seamless transition from $p$-values\newline
$\checkmark $ Easy to compute & \xmark $ $ Is no explicit Bayes' rule\newline
\xmark $ $ No theoretically justified thresholds are available\\
\hline
PD & Is the parameter $\theta$ positive or negative? & $\checkmark $ Confirmation\newline
    $\checkmark $ Rejection\newline
    of a hypothesis & $\checkmark $ Easy to interpret\newline
$\checkmark $ Easier to obtain than all other indices & \xmark $ $ Is no explicit Bayes' rule\newline
\xmark $ $ Limited usefulness as only the direction of an effect is captured\\
\hline
FBST and $e$-value & How large is the Bayesian evidence against $H_0$? & \xmark $ $ Confirmation\newline
    $\checkmark $ Rejection\newline
    of a hypothesis & $\checkmark $ Can be derived as an explicit Bayes' rule\newline
    $\checkmark $ Allows seamless transition from $p$-values while following the likelihood principle\newline
    $\checkmark $ Invariant under reparameterisations and is not troubled by nuisance parameters & \xmark $ $ Does not allow to confirm a hypothesis\\
\hline  
\end{tabular}
\label{tab:comparison}
\end{table*}

The first thing to note is that the various available posterior indices differ in their typical research question. For example, the Bayes factor $BF_{01}$ measures the relative change in beliefs towards $H_0$ relative to $H_1$ necessitated by observing the data $x$, while the PD answers the question if the parameter is of positive or negative sign (or equivalently, if an effect is positive or negative). The MAP-based $p$-value and the FBST and $e$-value both target the evidence against a null hypothesis $H_0$, while the ROPE can answer the question if the parameter of interest is practically equivalent to the values specified by the ROPE $R$. As a consequence, if the goal of a study is the rejection of a null hypothesis, the natural candidates are the FBST and $e$-value and the MAP-based $p$-value. However, as the FBST and $e$-value is theoretically much better justified we recommend to use the FBST in these situations. An alternative is given by the Bayes factor which can also reject a null hypothesis $H_0$ by indicating a strong necessity of a change in relative beliefs towards $H_0$.

If on the other hand, the goal is the confirmation of a research hypothesis $H_0$ or the confirmation of a theory, the Bayes factor and the ROPE are natural candidates. Both posterior indices allow for confirmation of $H_0$ and if the hypothesis is \textit{precise}, the Bayes factor should be preferred.

However, if there are reasons to assume that the existence of a precise effect is unrealistic, the ROPE may be more appropriate. Note that Stern \& Pereira \citeyear{Stern2020} remark concerning future research: ``In the context of information based medicine (...) it is important to compare and test the sensibility and specificity of alternative diagnostic tools, access the bio-equivalence of drugs coming from different suppliers (...)'' \cite[p.~9]{Stern2020}. While Kelter \citeyear{Kelter2020BayesianPosteriorIndices} recently investigated the specificity of various Bayesian posterior indices including the $e$-value, judging the bio-equivalence of drugs seems possible easily via the ROPE and HPD approach proposed by Kruschke \citeyear{Kruschke2018a}, see also the U.S. FDA industry guidelines for establishing bioequivalence \citeyear{USFoodAndDrugAdministrationBioequivalence2001}.

These recommendations offer guidance in determining default ROPEs, and in cases where no default ROPEs are well-established, it should be possible to determine a reasonable region of practical equivalence by incorporating available knowledge of prior studies and the measurement precision. For example, if similar research has yielded a specific range of parameter estimates, these can be used to determine a ROPE. Also, if measurements are taken with a specific precision (e.g. due to technological or biochemical properties) the ROPE can be determined based on which values can not be differentiated in practice due to finite measurement precision. Also, prior subject-domain knowledge may help in constructing a ROPE, for example, based on biochemical or physiological aspects which help in determining which parameter values can be treated as practically equivalent. For example, different blood pressure values are in practice separated only up to a specific precision.

Formally, the PD can also be used to confirm a hypothesis, but it is assumed that in most cases researchers are less interested in stating only the direction of an effect but much more in quantifying its size. As the PD is primarily targeted in answering the question of whether an effect's sign is positive or negative, we do not recommend it over the Bayes factor or ROPE.

The benefits and limitations listed in table 2 also should be considered when choosing an appropriate index in practice. Next to the research goal and study design which narrow down the set of suitable candidates, the interpretation, computation and theoretical properties should be taken into account.

For example, the Bayes factor, ROPE and FBST can be derived by minimisation of an appropriate loss function, making them a ``Bayesian'' procedure. In contrast, the PD and the MAP-based $p$-value are justified more heuristically. On the other hand, the interpretation of the PD and MAP-based $p$-value is much easier compared to the FBST or ROPE. Some indices like the FBST and $e$-value and the MAP-based $p$-value allow a seamless transition from NHST and $p$-values as they are designed to mimic $p$-values. Other indices require more sophisticated methodological shifts, like the Bayes factor or the ROPE.

Computational aspects also play a role: While all indices are in some form based on the posterior distribution visualised in figure \ref{fig:1}, the computation, in general, can become difficult. For example, advanced computational techniques like bridge sampling or the Savage Dickey density ratio may be necessary when no analytic solutions are available for the Bayes factor. In contrast, the FBST requires only numerical optimisation and integration (although this can become difficult in high-dimensional models, too). However, for most standard models used in psychology, all posterior indices discussed can be obtained with moderate computational effort \shortcite{Makowski2019a,Kelter2020BayesianPosteriorIndices}.

\section{Discussion}
Hypothesis testing stays an essential statistical method in psychology and the cognitive sciences. The debate about null hypothesis significance testing (NHST) goes on, and the existing problems are far from being solved \shortcite{Pashler2012a,OpenScienceFoundation,Matthews2017}. Among the proposed solutions to the replication problems in psychology caused by the inappropriate use of significance tests and $p$-values is Bayesian data analysis \shortcite{Wagenmakers2016,VanDongen2019}. However, practical Bayesian hypothesis testing in psychology is challenged by the availability of various posterior indices for significance and the size of an effect. This complicates Bayesian hypothesis testing because the availability of multiple Bayesian alternatives to the traditional $p$-value causes confusion which one to select in practice and why. 

In this paper, various Bayesian posterior indices which have been proposed in the literature were compared and their benefits and limitations were discussed. Also, guidance on how to select between different Bayesian posterior indices for hypothesis testing in practice was provided.

To conclude, redirect attention to the running example. Based on the running example, the Bayes factor was indecisive. The ROPE did neither reject the null hypothesis nor accept it, although it only slightly failed to reject the null hypothesis. The MAP-based $p$-value also did not reject the null hypothesis. The probability of direction expressed strong evidence that the null hypothesis $H_0:\delta=0$ is not true. The FBST also expressed strong evidence against the null hypothesis $H_0:\delta=0$. However, both the PD and FBST were not \textit{certain} (and for the FBST it would also have been possible to use the distribution of the $e$-value instead). How should one interpret the differences obtained by employing the various Bayesian posterior indices? One option is to hand off this problem to statisticians and to prompt them to agree on one single measure to use in practice. This is probably not going to happen anytime soon, so the second (and more appealing) option is to incorporate the study design and research goal into the decision which index to use.

In the running example, a Bayesian two-sample t-test was carried out and it was intentionally not stated (1) what kind of data was measured and (2) what the existence of an ``effect'' describes. For example, if data observed in both groups is the blood pressure which is measured via a standard procedure up to a specific precision, an effect may be described as the difference in average blood pressure between a treatment group taking a drug for lowering blood pressure and a control group. In such a setting, it seems unrealistic to assume the existence of an \textit{exact} null effect $\delta=0$ (even if the drug works, some difference between both groups is to be expected). It is more appropriate to employ the ROPE $R=[-0.1,0.1]$ to test if the effect size is \textit{practically equivalent to zero}.

Consider a different study design in which the participants in both groups are partnered. Data measured in the first group is the height in inches of each study participant. Each participant in the second group now guesses the height in inches of her partner without seeing her. The null hypothesis $H_0:\delta=0$ states that participants in the second group have the ability of extrasensory perception and can perfectly determine their partner's height without having seen them previously. In this case, a \textit{precise} null hypothesis $H_0:\delta=0$ is reasonable and when the goal is a rejection of this hypothesis, the FBST or MAP-based $p$-value could be used. If in contrast, the goal is the confirmation (or the goal is not specified as confirmation or rejection), the Bayes factor could be used.

The differences obtained by using the various available posterior indices for the same running example are therefore no contradictions to each other, but much more the results of the different assumptions each method makes. The suitability of a given index to a study or experiment depends on both on the experimental design and the research goal.

Notice that we do not advocate against or in favour of the general use of one of the available indices. However, we want to draw attention to two posterior indices which have been widely underutilised in psychological research.

First, in a variety of situations, the ROPE seems to be appropriate by not making the often unrealistic assumption of an exact effect. This property is appealing in particular in psychological research, as often some kind of effect is expected to be observed only due to the randomness and noise in the experimental data. Also, for effect sizes and regression coefficients there exist plausible default values and the procedure can be identified as a formal Bayes' rule.

Second, the FBST is an appealing option when transitioning from NHST and $p$-values to Bayesian statistics. To our knowledge, the FBST is still widely underused in psychological practice. Also, the ROPE is still widely underused, although the situation is a little better than for the FBST.

This situation may be attributed to the more statistical background of the FBST and the relatively new proposal of the ROPE, but also the lack of easy-to-apply software implementations \shortcite{Makowski2019a,Jasp2019,VanDoorn2019,Kelter2020,Kelter2020JORSBayest}. However, applying any of the discussed indices in this paper is straightforward as shown in the OSF supplemental file, and we encourage readers to reproduce all analyses and results. 

Based on the recommendations given, readers can decide themselves which index is most useful and makes the most sense in their study or experiment. The guidance provided here also shows that future research needs to be conducted which analyses how the various indices relate to each other both theoretically and in practice \shortcite{Makowski2019,Kelter2020BayesianPosteriorIndices}.

In summary, this paper hopefully guides how to select between different Bayesian posterior indices for hypothesis testing in psychology and fosters critical discussion and reflection when considering hypothesis testing in the Bayesian.

\section{Author Contributions}
R. Kelter is the sole author of this article and is responsible for its content.


\section{ORCID iD}
Riko Kelter \newline \url{http://orcid.org/0000-0001-9068-5696}

\section{Declaration of Conflicting Interests}
The author(s) declare that there were no conflicts of interest with respect to the authorship or the publication of this article.

\section{Open Practices}
Open Data: See \url{https://osf.io/xnfb2/}\newline
Open Materials: See \url{https://osf.io/xnfb2/}\newline
Preregistration: not applicable

\bibliographystyle{mslapa}
\bibliography{library}
\newpage

\end{document}